\documentclass[
    ,final            
  ]
  {aipproc}

\layoutstyle{8x11double}

\begin{document}

\title{Low Frequency Observations of Millisecond Pulsars with the WSRT}


\author{B. W. Stappers}{
  address={Stichting ASTRON, Postbus 2, 7990 AA Dwingeloo, The Netherlands},
  altaddress={Astronomical Institute ``Anton Pannekoek'', University of Amsterdam, Kruislaan 403, 1098 SJ Amsterdam, The Netherlands}
}

\author{R. Karuppusamy}{
  address={Stichting ASTRON, Postbus 2, 7990 AA Dwingeloo, The Netherlands},
  altaddress={Astronomical Institute ``Anton Pannekoek'', University of Amsterdam, Kruislaan 403, 1098 SJ Amsterdam, The Netherlands}
}
\author{J. W. T. Hessels}{
  address={Astronomical Institute ``Anton Pannekoek'', University of Amsterdam, Kruislaan 403, 1098 SJ Amsterdam, The Netherlands}
}

\begin{abstract}
With LOFAR beginning operation in 2008 there is huge potential for
studying pulsars with high signal to noise at low frequencies. We
present results of observations made with the Westerbork Synthesis
Radio Telescope to revisit, with modern technology, this frequency
range. Coherently dedispersed profiles of millisecond pulsars obtained
simultaneously between 115-175 MHz are presented. We consider the
detections and non-detections of 14 MSPs in light of previous
observations and the fluxes, dispersion measures and spectral indices
of these pulsars.  The excellent prospects for LOFAR finding new MSPs
and studying the existing systems are then discussed in light of these
results.
\end{abstract}


\maketitle


\section{Introduction}

There are a large number of new radio facilities currently in the
planning or construction phase and it is important that we consider
the impact of these telescopes on all aspects of pulsar research. The
first of these facilities to come on line will be the low frequency
telescopes like LOFAR, LWA and MWA, which from now on we will
collectively refer to as low frequency radio arrays (LRAs). While
these instruments differ somewhat in design and frequency range, they
all work at frequencies below 300 MHz and will be used to study the
existing pulsar population and to discover new pulsars. It is
therefore appropriate to consider what we might expect from pulsars in
this frequency range. As discussed elsewhere in this volume by van
Leeuwen \& Stappers \ref{} there is huge potential for finding new
pulsars with LOFAR for example. Their study didn't consider the
millisecond pulsars (MSPs) as less is presently known about their
low-frequency properties. The main issues governing the potential the
LRAs have for discovering new MSPs are; the steepness of the radio
spectrum and whether it turns over in this frequency range and the
magnitue of the scattering in the interstellar medium.

In the late nineties and earlier this decade Kuzmin \& Losovsky
\cite{kl99,kl01} published the first papers which took a statistical look at the
MSPs in the frequency range of interest here. They first presented a
number of pulse profiles at frequencies near 100 MHz and by comparing
them with profiles at higher frequencies they concluded that unlike
the normal radio pulsar population there was little evidence for
broadening of the pulse profile as a function of frequency
\cite{kl99}. They then went on to show that the radio spectral index of the
MSPs seemed not to show a turnover at frequencies at or near 120 MHz
as do the majority of higher magnetic field pulsars
\cite{kl01}. This later result, combined with the relatively large 
number of MSPs they detected, indicated that  the low
frequency range could potentially be a valuable one for finding MSPs.

\begin{table}
\begin{tabular}{lrrrrcc}
\hline
\tablehead{1}{c}{b}{Pulsar} &
\tablehead{1}{c}{b}{Period (ms)} &
\tablehead{1}{c}{b}{DM (cm$^{-3}$ pc)} &
\tablehead{1}{c}{b}{Flux (mJy)} &
\tablehead{1}{c}{b}{Flux Error (mJy)} &
\tablehead{1}{c}{b}{Profile} &
\tablehead{1}{c}{b}{WSRT Detection} \\
\hline
J0034-0534  &	1.87  	&13.7	&250	&120     & y	  &    y    \\
J0218+4232 	&2.33  &	61.2	&270	&150      &n	 &     n    \\
J0613-0200 	&3.06  &	38.8	&240	&100      &n	 &     n    \\
J0621+1002 	&28.85 &	36.6	&50	&25       &n	  &    n    \\
J1012+5307 &	5.25  &	9.0	&30	&15       &y	      &y    \\
J1022+1001 &	16.45 &	10.2	&90	&40       &y	      &y    \\
J1024-0719 &	5.16  &	6.4	&200	&100      &y	      &n    \\
B1257+12   &	6.21  &	10.1	&150	&50       &y	      &y    \\
J1713+0747 &	4.57  &	15.9	&250	&100      &y	      &n    \\
J1744-1134 &	4.09  &	3.1	&220	&100      &n          &  y    \\    
J1911-1114 &	3.62  &	30.9	&260	&130      &y	      &y    \\
J2051-0827 &	4.51  &	20.7	&250	&100      &n	      &n    \\
J2145-0750 &	16.05 &	9.0	&480	&120      &n	      &y    \\
B1957+20   &     1.61 &   29.11 &       &         &           &       y\\
\hline    
\end{tabular}
\label{stappers:results-table}
\caption{Pulsars observed so far with the LFFEs at the WSRT. The sixth column indicates whether a 100-MHz profile is presented either in \cite{kl99} or on the EPN database. PSR B1957+20 was not previously observed at these frequencies.}
\end{table}

Kuzmin \& Losovsky had to use very narrow bandwidths (32 $\times$ 5
kHz) to obtain their results due to the deleterious effects of
dispersion in the interstellar medium and they had limited time
resolution (at best 0.64 -- 0.128 ms). We therefore decided to obtain
higher resolution pulse profiles from a number of these MSPs in order
to better determine the pulse profile changes and the influence of
scattering. We first discuss the observations and present our
results. We then discuss the implications for the properties of MSPs
and scattering in the interstellar medium and for their future study
and detection with the LRAs.

\section{Observations and Data Reduction}

We observed a total of 14 pulsars (see Table
\ref{stappers:results-table}) using the low frequency front ends
(LFFEs) on the Westerbork Synthesis Radio Telescope (WSRT). The LFFEs
have good sensitivity in the frequency range 115-180 MHz, where the
lower limit is defined by the FM band and the upper limit by the
response of the feeds. The band does contain some interference which
is especially troublesome because the data is only sampled with 2
bits. We therefore selected eight clean bands each of 2.5 MHz
bandwidth distributed throughout the band at 116.75, 130, 139.75,
142.25, 147.5, 156, 163.5 and 173.75 MHz. The data were oversampled at
40 MHz, decimated in real-time to 2.5 MHz bandwidth and then baseband
recorded using the PuMa II pulsar backend.

The data were reduced using the open source software package, DSP for
Pulsars, DSPSR\footnote{http://sourceforge.net/projects/dspsr/}. A
coherent filterbank of either 32 or 64 channels in each of the 8 bands
was formed offline. The data were coherently dedispersed in each of
the 32 or 64 channels, leading to a final time resolution of 25.6 or
51.2 $\mu$s. The data was also folded offline with an average pulse
profile being formed for every ten seconds of data. These
time-frequency cubes for each ten seconds of data were then checked
for interference using a median filtering technique based on the rms
noise in each frequency channel. The cleaned ten-second average
profiles from all 8 bands were then summed to form a profile for each
band and these bands were subsequently combined. It soon became
evident that the combination of a wide range of frequencies and the
low central frequencies meant a very accurate, epoch specific,
determination of the dispersion measure (DM) was required in order to
properly combine the 8 bands. Each data set was therefore optimised
for the best DM, by maximising the signal-to-noise ratio of the pulse
profile, and the inter-channel and inter-band dispersion correction
was redone with the new best DM value. Typically the DM values needed
to be changed at the $10^{-3} - 10^{-4}$ level, indicating the
importance of accurate DM determinations for these low frequency
observations. It will be interesting to see if follow-up observations
at these frequencies also indicate small differences in the DM and
thus a small DM optimisation step will be required for any analysis at
these frequencies. The potential for using these accurate
determinations of the DM for other applications like high precision
pulsar timing at higher frequencies needs to be investigated.

\section{Results and Discussion}

\begin{figure}[htb]
  \includegraphics[height=\linewidth,angle=-90]{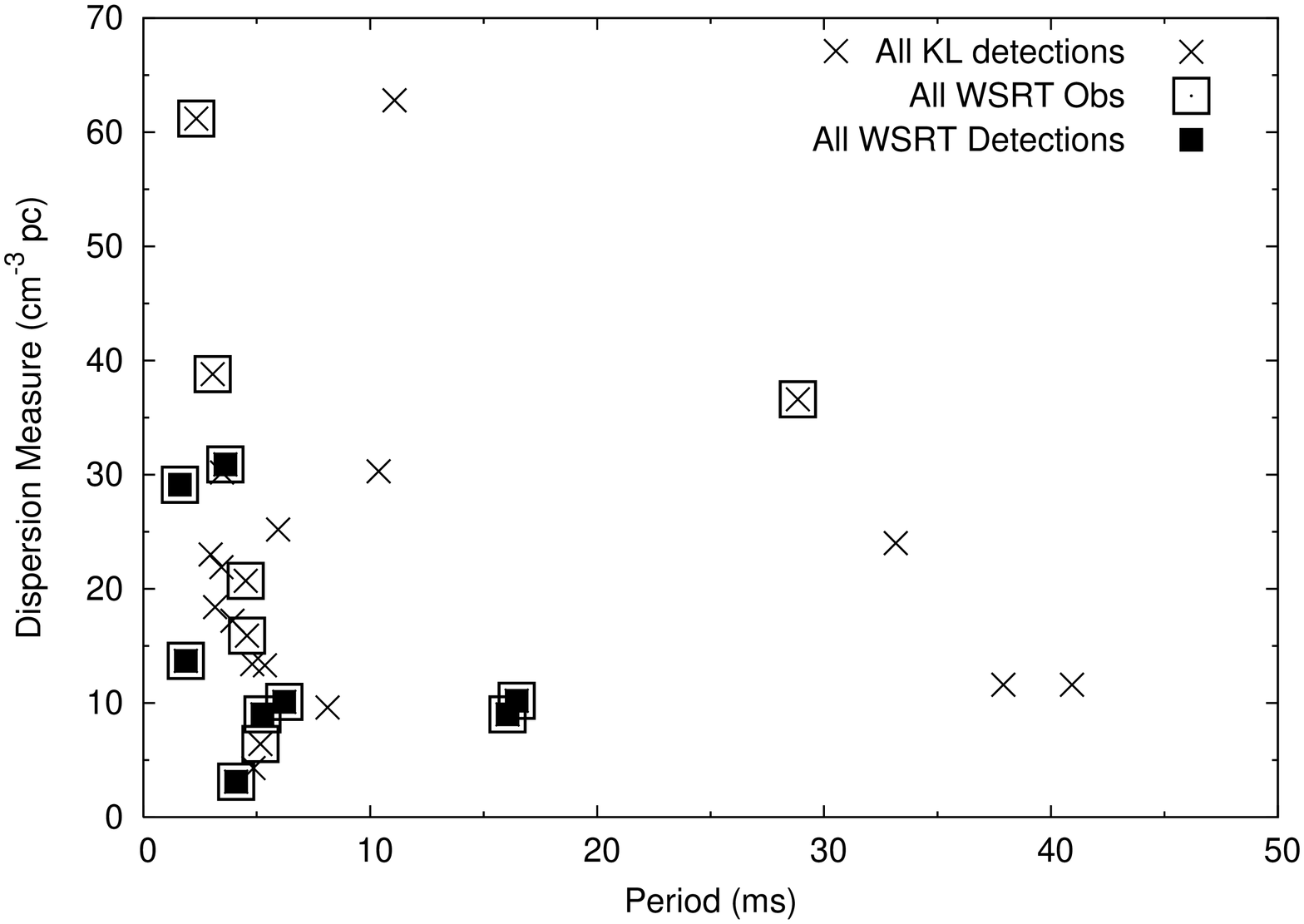}
  \includegraphics[height=\linewidth,angle=-90]{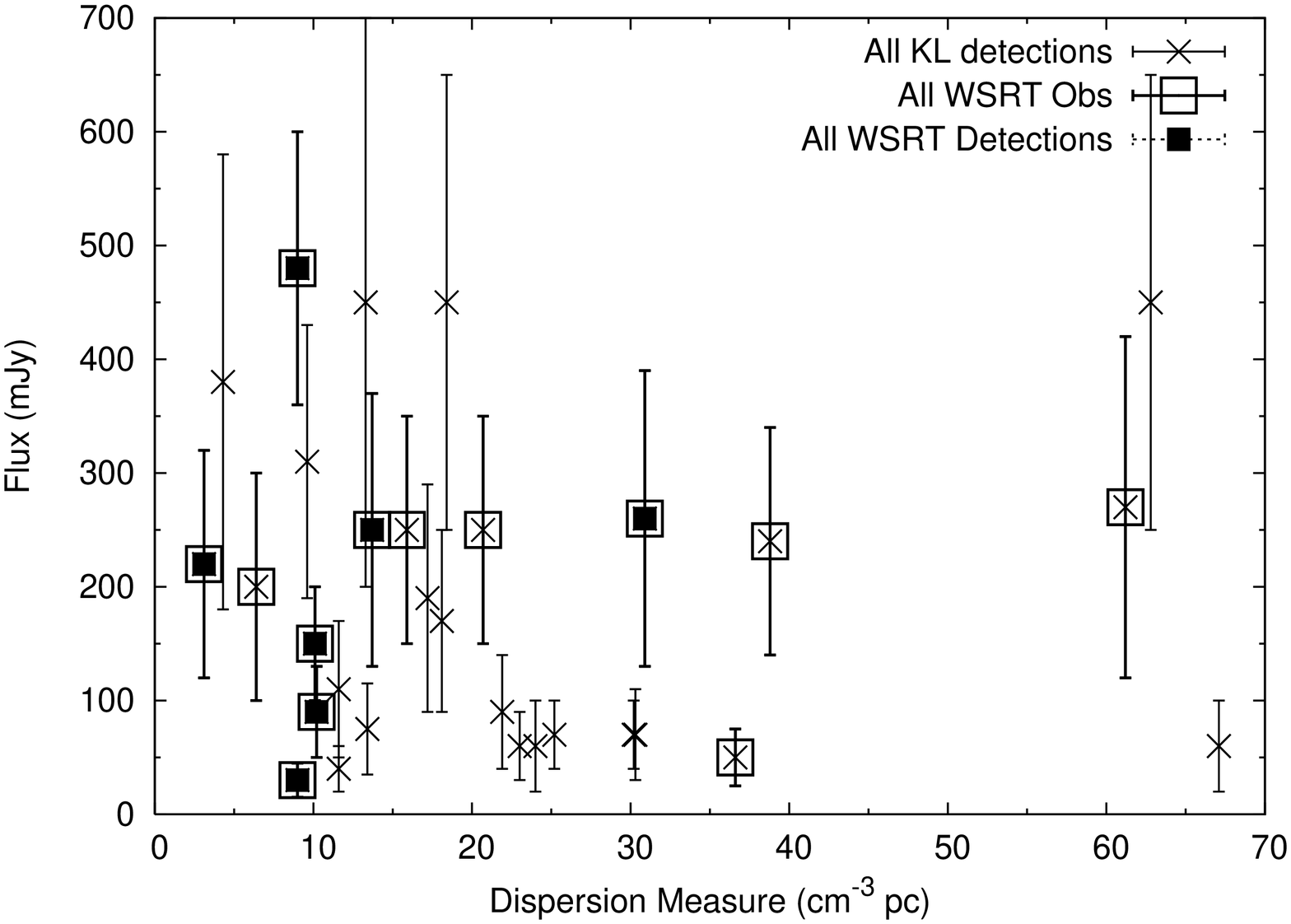}
  \label{stappers:kuzmincf-figure} \caption{Plots comparing detections
  made with the LFFEs on the WSRT with the detections presented in
  KL99 and KL01.}
\end{figure}

We present the results of the observations with the LFFEs in Table
\ref{stappers:results-table}. Of the 14 pulsars observed a clear 
detection of a pulse profile was made in 8 cases and of the 13 pulsars
which overlapped with the sample of KL99/KL01 7 were detected. In
Figure \ref{stappers:kuzmincf-figure} the detections are plotted as a
function of DM and flux in an attempt to determine
whether there is a common reason for detection or non-detection. All
the sources detected by KL99/KL01 are plotted with crosses, while
those observed by us are indicated by the open squares. The closed
squares correspond to our detections. Before considering the sources
as a whole we'll discuss a couple of individual cases. 

The pulsar with the worst period and DM combination and thus the one
that is most likely to suffer from scattering is PSR J0218+4232
(top-leftmost point in the left hand plot of figure
\ref{stappers:kuzmincf-figure}). Despite this KL01 claim to have
detected it. It is unclear whether they detected pulsations as only a
flux is quoted in their paper and it is well known that it is a bright
point source all the way down to 30 MHz. However we do not detect it
in our observations and also do not detect it in deep observations at
250 MHz. This suggests that this source is not seen at these
frequencies due to scattering.

PSR B1957+21, like PSR J0218+4232, has one of the worst combinations
of period and DM and yet we detected it with very high
signal-to-noise. This is the only pulsar in our sample that was not
observed by KL99/KL01 and this is probably because they had
insufficient time resolution. Not only did we detect the source but
the signal-to-noise ratio was sufficiently high that it was detected
in every 10-second interval in each of the eight bands.  Moreover the
observations took place just as the pulsar was coming out of
eclipse. The wide fractional frequency range simultaneously spanned by
these data will provide an exciting opportunity to study the
properties of the eclipses in this system.

It is apparent from Figure \ref{stappers:kuzmincf-figure} that there
is no clear relationship between detection of a source in our
observations with any combination of period and DM nor with the quoted
100 MHz flux. However the majority of the non-detections are at the
lower fluxes, although not all. Seven pulsars in our sample have
published profiles either in KL99 or in the EPN
database\footnote{http://www.mpifr-bonn.mpg.de/div/pulsar/data/} and
of those we detect five. It remains unclear why the other sources were
not detected and if all the detections by KL99/KL01 are secure then it
points to some time variable phenomenon.

It is unlikely to be due to scintillation, as the scintillation
bandwidth decreases rapidly with frequency and the relatively wide
bands used here mean that there are many scintles in each band.  The
average flux should therefore remain relatively constant.  At these
frequencies interference is always a concern and it is certainly
variable on the timescales of our observations. However all of the
sources we detected were seen in the individual 2.5 MHz bands and
these are widely separated in frequency and therefore one would not
expect them all to be affected by interference
simultaneously. Moreover inspection of the data where pulsars were not
seen does not show overly worse interference conditions than when
pulsars were detected.

As discussed above, for the pulsars that we have detected, corrections
had to be made to the DM in order not to have a broadened
profile. Changes in DM are therefore another variable that might
affect our ability to detect the MSPs at low frequencies. However, in
order for the profile to be smeared by a large fraction of the pulse
period across the 2.5 MHz band of each observing band then DM changes
of the order of 0.1 are required. This is two to three orders of
magnitude more than were detected above and is much larger than has
been measured for any MSP \cite{yhc+07}. It is therefore unlikely that
this is the reason for the lack of detection of some of these sources.

While not thought to be highly variable, one of the main reasons why
one might not expect to be able to detect some MSPs at these
frequencies is scattering in the interstellar medium. The combination
of the short rotational periods and the extreme frequency dependence
of scattering mean that the pulse profiles may be scattered by more
than a pulse period and thus it is no longer possible to detect them
as pulsed sources. While there does appear to be an empirical relation
between the DM and scattering (e.g. \cite{cl01,bcc+04}) the more than
a couple of orders of magnitude variation about the relation means
that it has litle predictive power (e.g. see Figure 4 of
\cite{bcc+04}). This, combined with the results of KL99/KL01 and this
work show that it is basically only possible to determine which MSPs
will be visible by actually observing them.

\begin{figure}
  \includegraphics[height=\linewidth,angle=-90]{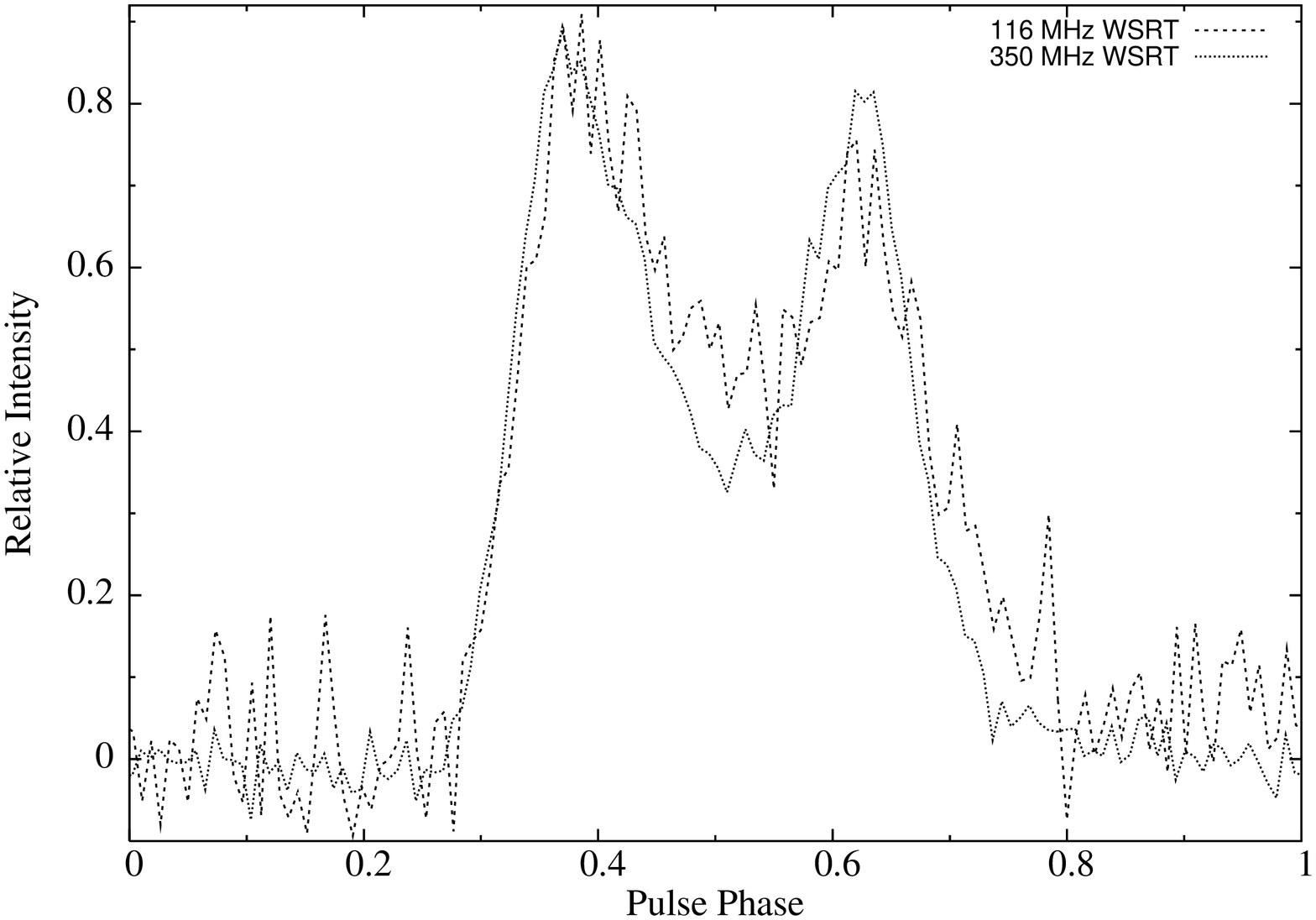}
  \includegraphics[height=\linewidth,angle=-90]{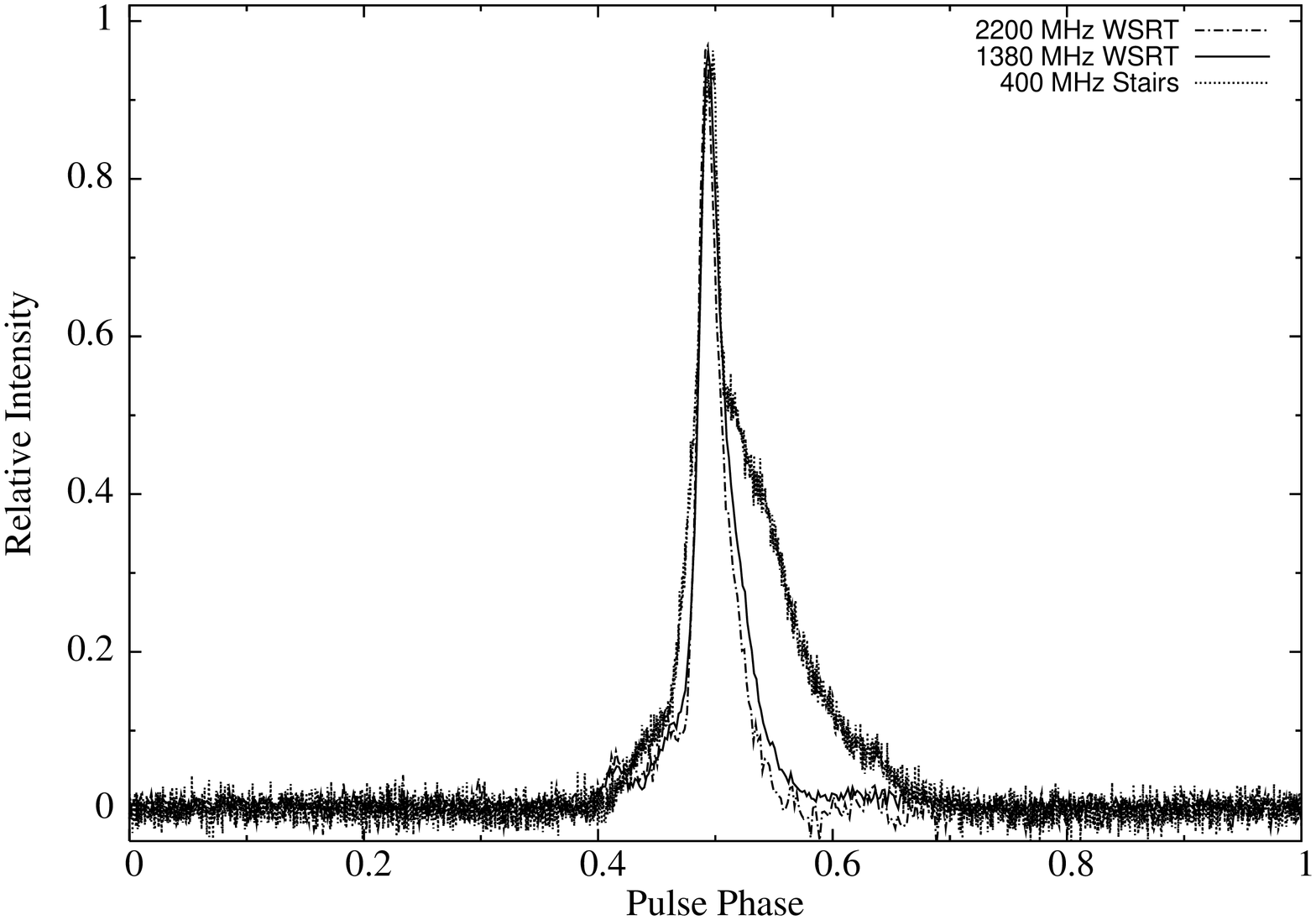}
  \caption{A comparison of the pulse profile evolution of PSR
  J0034-0534 (left) and PSR J1713+0737 (right). Both pulsars have
  similar DMs and the expected fluxes in the LFFE band
  are similar, however PSR J1713+0747 is not detected. There is some
  broadening of the profile of PSR J1713+0747 however it is not clear
  if this is due to scattering or profile evolution. The 400 MHz
  profile is from Arecibo data and is provided by I. H. Stairs.}
\label{stappers:0034-1713}
\end{figure}

\section{PSR J0034-0534 and PSR J1713+0737}

PSRs J0034-0534 and J1713+0737 are a pair of pulsars which illustrate
the unpredictability of the degree of scattering. Both pulsars are
claimed to be detected by KL99/KL01 and the pulse profiles can be
found in the EPN database. A simple comparison of the two profiles
suggests that they were both detected equally well, apart from the
time resolution being better for PSR J1713+0747. The fluxes quoted for
the two sources at these frequencies are also very similar. However,
we easily detect PSR J0034-0534 and do not see PSR J1713+0747 at all.

What could be the reason why we see one of these pulsars and not the
other? As well as the fluxes being similar they also have very similar
DMs, however we have already discussed the fact that there is not a
very robust correlation between DM and the degree of
scattering and so it may be that the profile of PSR J1713+0747 is too
scattered to be detected (assuming also, with no explicit reason, that
the detection by KL99 is not real).  To test this we plot in Figure
\ref{stappers:0034-1713} the frequency evolution of the
average pulse profiles of the two pulsars. The comparison between the
328 and 116 MHz pulse profiles of PSR J0034-0534 shows a small degree
of broadening of the pulse profile due to scattering however it is
still clearly detected at the lowest frequencies. The situation is a
little bit more complicated for PSR J1713+0747 where at the higher
frequencies there is some evidence for a broadening of the pulse
profile but there is also, what appears to be, the development of a
new component on the trailing edge of the profile at 400~MHz. It is
therefore unclear whether scattering is the reason why the pulsar is
undetected at these frequencies. 

\section{The pulse profile of PSR J2145-0750}

PSR J2145-0750 is presently the brightest MSP observed at frequencies
below 200 MHz. Kuzmin \& Losovsky (1996; KL96)\nocite{kl96} first
detected the source near 100 MHz and they compared their observed
profile with those obtained at higher frequencies. As a result of
Gaussian fitting to the profiles at 102, 430 and 1520 MHz they find
that the profile apparently broadens at higher frequencies. This is
extremely unusual as the traditional view is that the profiles become
narrower at higher observing frequencies and this is thought to
indicate that the higher frequency emission comes from deeper in the,
predominantly dipolar, magnetosphere. The authors then suggest that
this may be interpreted as being evidence for a magnetic field where
the quadrupole terms are also important. Multipole contributions might
be expected in MSPs because they have much more compact magnetospheres
and this was some of the first evidence that this might be observable.

\begin{figure}
  \includegraphics[height=\linewidth,angle=-90]{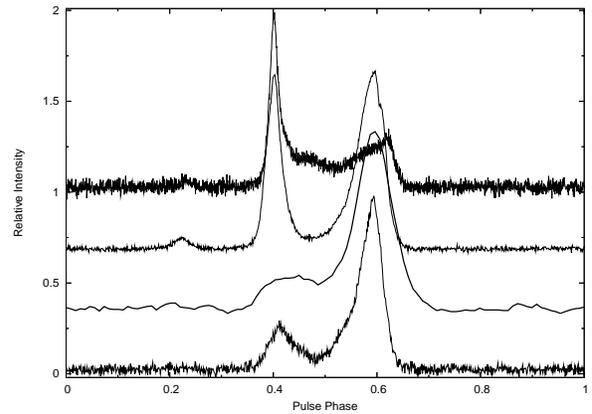}
  \caption{The profile evolution of PSR J2145-0750 as a function of
  frequency is shown. The topmost plot corresponds to a frequency of
  1380 MHz, the second is 350 MHz, the third is the KL99 plot at 100
  MHz and the lowest profile is the WSRT detection at 150 MHz.}
\label{stappers:2145}
\end{figure}

In Figure \ref{stappers:2145} we present observations made at the WSRT
using the PuMa II backend at frequencies of 150, 350 and 1380 MHz. In
all cases the data were coherently dedispersed and the time resolution
was at worst 25.6 $\mu$s. Also shown is the 102 MHz profile from
KL96. One can see the advantages of the coherent dedispersion in our
profile at 150 MHz which is significantly sharper. It is clear that
the leading component (component I in KL96) has undergone significant
frequency evolution and also that the the peak separation between it
and the second large peak is reduced. However the shift is smaller
than seen by KL96 as their profile seems to be slightly distorted,
perhaps due to dispersive smearing. Moreover it is not clear that this
is still component I as it has such a different shape. It now more
resembles the trailing component of the high frequency profile. A
multipole interpretation would require further observations at smaller
frequency intervals to better track the component evolution.

\section{Conclusions}

We have used the LFFEs on the WSRT to make low frequency observations
of 14 MSPs and successfully detected 8 of them. Using the PuMa II
backend and coherent dedispersion we are able to get profiles with
significantly higher time resolution than was previously
possible. Previous work has suggested that we should have been able to
detect the majority of these sources. We consider whether our failure
to detect them is due to scattering or flux limits and find no clear
factor which governs our ability to them. Taking the claimed
detections in the literature at face value it would therefore suggest
that there is some time dependent effect which is lowering the flux of
them.

We consider diffractive scintillation, interference, and DM variations
and find that neither can plausibly explain the non-detections. One
possible explanation is refractive scintillation which causes flux
modulation on long timescales. However further investigation into the
long term flux stability of the sources would be required to confirm
this effect as the modulation due to refractive scintillation is
expected to be low \cite{ssh+00}.

What do these observations tell us about the prospects for observing
and detecting MSPs with the LRAs? For LOFAR the sensitivity in the
frequency range discussed above is expected to be at least 20 times
better than the WSRT-LFFE combination. This means that we can expect
to have the sensitivity to discover new MSPs with LOFAR. We need better
statistics before being able to determine any sort of MSP luminosity
function in this frequency range, but the initial results are very
promising. It is also apparent that it will not be possible to
determine a priori the DM out to which MSPs might be detected. The huge
spread around the DM-scattering relationship precludes that and so a
search out to DMs up to at least 100 will be a necessary component of
searches for MSPs.

The greatly improved sensitivity of the LRAs over the existing
telescopes, in general, means that they will also be able to study the
single pulses from a large sample of MSPs for the first time. This
will be essential for determining whether there are any changes in the
single pulse properties of MSPs, with their significantly smaller
magnetic fields, compared to the normal pulsar population. That is to
say, do any of the known single pulse properties depend on rotation
rate or magnetic field strength or even neutron star surface
temperature.

The LRAs have the potential to not only increase the number of MSPs
known but also to study their emission properties with unprecedented
detail.


\begin{theacknowledgments}
We would like to thank the staff of the WSRT for assistance with
obtaining the data used in this paper. J.W.T.H. thanks NSERC and the
Candian Space Agency for a postdoctoral fellowship and supplement
respectively.
\end{theacknowledgments}


\bibliographystyle{aipproc}   



\begin{thebibliography}{7}
\expandafter\ifx\csname natexlab\endcsname\relax\def\natexlab#1{#1}\fi
\providecommand{\enquote}[1]{``#1''}
\expandafter\ifx\csname url\endcsname\relax
  \def\url#1{\texttt{#1}}\fi
\expandafter\ifx\csname urlprefix\endcsname\relax\def\urlprefix{URL }\fi
\providecommand{\eprint}[2][]{\url{#2}}

\bibitem[{Kassim} and {Lazio}(1999)]{kl99}
N.~E. {Kassim}, and T.~J.~W. {Lazio}, \emph{Astrophys.\,J.} \textbf{527},
  L101--L104 (1999).

\bibitem[{Kuzmin} and {Losovsky}(2001)]{kl01}
A.~D. {Kuzmin}, and B.~Y. {Losovsky}, \emph{aap} \textbf{368}, 230--238 (2001).

\bibitem[{You} et~al.(2007)]{yhc+07}
X.~P. {You}, G.~{Hobbs}, W.~A. {Coles}, R.~N. {Manchester}, R.~{Edwards},
  M.~{Bailes}, J.~{Sarkissian}, J.~P.~W. {Verbiest}, W.~{van Straten},
  A.~{Hotan}, S.~{Ord}, F.~{Jenet}, N.~D.~R. {Bhat}, and A.~{Teoh},
  \emph{mnras} \textbf{378}, 493--506 (2007), \eprint{arXiv:astro-ph/0702366}.

\bibitem[{Cordes} and {Lazio}(2001)]{cl01}
J.~M. {Cordes}, and T.~J.~W. {Lazio}, \emph{Astrophys.\,J.} \textbf{549},
  997--1010 (2001).

\bibitem[{Bhat} et~al.(2004)]{bcc+04}
N.~D.~R. {Bhat}, J.~M. {Cordes}, F.~{Camilo}, D.~J. {Nice}, and D.~R.
  {Lorimer}, \emph{Astrophys.\,J.} \textbf{605}, 759--783 (2004).

\bibitem[Kuzmin and Losovskii(1996)]{kl96}
A.~D. Kuzmin, and B.~Y. Losovskii, \emph{Astr.\,Astrophys.} \textbf{308},
  91--96 (1996).

\bibitem[Stinebring et~al.(2000)]{ssh+00}
D.~R. Stinebring, T.~V. Smirnova, T.~H. Hankins, J.~Hovis, V.~Kaspi,
  J.~Kempner, E.~Meyers, and D.~J. Nice, \emph{Astrophys.\,J.} \textbf{539},
  300--316 (2000).

\end{thebibliography}


\end{document}